# Sliding and translational diffusion of molecular phases confined into nanotubes

Rémi Busselez, Claude Ecolivet, Régis Guégan, Ronan Lefort, Denis Morineau, Bertrand Toudic

Institute of Physics of Rennes, GMCM UMR-CNRS 6626
University of Rennes 1
F-35042, Rennes, France
E-mail: denis.morineau@univ-rennes1.fr

Mohammed Guendouz

Laboratoire d'Optronique, FOTON, CNRS-UMR 6082,
University of Rennes 1
F-22302 Lannion, France

Frédéric Affouard

LDSMM
University of Lille 1
F-59655 Villeneuve d'Ascq, France

**Abstract:** The remaining dynamical degrees of freedom of molecular fluids confined into capillaries of nano to sub-nanometer diameter are of fundamental relevance for future developments in the field of nanofluidics. These properties cannot be simply deduced from the bulk one since the derivation of macroscopic hydrodynamics most usually breaks down in nanoporous channels and additional effects have to be considered. In the present contribution, we review some general phenomena, which are expected to occur when manipulating fluids under confinement and ultraconfinement conditions.

# Introduction

The scientific interest of research on molecular fluids confined in nanoscopic cavities has been recently amplified by the emergence of nanosciences. Indeed, many crucial issues of this field, such as nanosensors, nanofluidic networks, or lab-on-chips, rely on the technological control of complex fluid phases down to the scale of a few molecules, at steady-state or under flow. However, the comprehensive understanding of the physical properties of nanomaterials is far from being thoroughly achieved, as new fundamental questions have been raised, challenging interdisciplinary domains of chemical physics including statistical physics, kinetics of out of equilibrium states, or collective dynamics.

At the micrometer scale, the properties of a phase are mainly governed by bulk quantities, which result from the average over a very large number of molecules. This is not valid anymore when reducing the size of the phase down to the nanometer scale. Indeed new effects are expected from the overlap and competition between the typical correlation lengths driving the phase structure or dynamics and the finite size of the system. Moreover, these boundaries define an interface, with another fluid or another physical state in the case of nanoparticles or nucleation of nanocrystallites, or with a solid matrix in case of nanoconfined fluids. The presence of this interface adds an external field that can profoundly modify the symmetry of the system, and adds a significant interfacial contribution to the total energy of the system. As a consequence, new equilibrium states of the condensed phase can be observed and strongly altered collective dynamical and relaxation properties are also expected.

The ultimate conceptual step of size reduction of a phase is reached by supramolecular self-assemblies, which realize the ultraconfinement of interacting species. It can be encountered in molecular clathrates or intergrowth compounds, defined as a potentially incommensurate organization of two or more molecular networks. In that case, the interfacial interaction between different constituents becomes so prevailing that collective structural or dynamical properties of the guest network and host counterpart are closely interrelated and cannot be disentangled. This distinction between host and guest is indeed irrelevant to advanced theoretical tools such as crystallography in spaces of dimension higher than three. From this point of view, the ultraconfinement realized at the nanoscale in such supramolecular edifices paradoxally reintroduces the occurrence of long-range orders in the material considered as a whole. They induce collective features that are reminiscing of bulk solid phases properties, like superspace groups, anomalous compressibility, or sliding modes dispersion.

# 1   Nanoconfinement of simple liquids

Because it reaches nanometer dimensions, mesoporous confinement induces new properties of the fluid, which are not simply deduced from the bulk one. Most of already addressed issues concern the phase behavior and especially the freezing/melting [1]. Another very active topic concerns the molecular dynamics of confined fluids and its relation with the glass transition [2].

Two major effects are usually invoked in mesoporous confinement. The first is cut-off or finite size effect. It states that neither static nor dynamical correlation length can grow larger than the maximum pore size. The second one is surface effect, which is introduced by the large surface-to-volume ratio encountered in system of nanometer-size [3,4].

Considering these two effects, one expects to reach a proper basic insight into the phenomenology of confined simple fluids. An improved tendency towards supercooling is achieved in small volumes, which reduce the probability of crystal nucleation. A simple consequence of finite size effects is the observed restriction of translational orders, which results in a broadening of the crystalline Bragg peaks and excluded volume effects for the structure factor of liquids. Surface interaction adds to the volume energy of the system. This parameter is taken into account in the Gibbs-Thomson theory for the melting point variation in the limit of a macroscopic thermodynamic approach. Considering the importance of fluid-wall interactions with respect to fluid-fluid ones allows one to envisage a large amorphous interfacial component in nanoconfined crystals or, for special substrates, the occurrence of unusual ordered contact layers [5]. Additional effects, such as pore dimensionality, should not be ignored in some special cases for instance for long alkanes solid phases or for marginally anisotropic confined liquid alcohols [6,7].

The molecular dynamics of confined liquids is certainly more complex than its thermodynamics is. Nonetheless, there are currently a growing number of evidences that interfacial interactions effects markedly dominate the structural relaxation of globular molecular liquids. The novel dynamical boundary condition introduced by the pore surface generally slows down the molecular dynamics by some orders of magnitude. The large increase of the local structural relaxation time is transmitted to the inner fluid through dynamical molecular correlations. This leads to the globally slower and strongly inhomogeneous dynamics of the confined fluid. A crucial issue is the nature of the correlation length associated with this surface-induced spatially heterogeneous dynamics. For simple systems, it is indeed commonly indentified to the cooperativity length of the relaxation dynamics of the confined liquid [8]. Variable diameter confinement experiments have been thought as a possible measurement of the cooperativity length associated to the glassy dynamics of supercooled liquids. This supposes of course that the thermodynamical state of the bulk liquid is maintained in confinement, which does not seem to be the usual case [7]. For instance, the liquid density and the molecular arrangement markedly differ from the bulk ones for systems sizes of a few nanometers [9]. Eventually, finite size has to be considered as the other dominant effect that may affect the dynamics of supercooled liquids. Most likely, the only modest increase of the cooperativity length in supercooled liquids requires very small confinement diameter for finite size effects to be effective, which precludes surface effects and thermodynamical state of the confined phase to be fully controlled.

## 2  Complex fluids confined in mesoporous channels

However, it rapidly appears that interpretations based only on cut-off and interfacial effects break down when considering more complex fluids.
Molecular fluids relevant for technological applications based on nanoconfinement (fluidics, optronics, biotechnology…) are usually complex systems including soft matter and bio-solutions. They display additional features related to molecular self-assembling, various phase transitions and multiple relaxation processes. Additional effects introduced by the fluid manipulation at the nanoscale are to be considered, such as topological aspects of the confinement that can become essential when anisotropic parameters of the fluid appear relevant. In some cases, orientational or translational order parameters can be driven by quenched disorder, low dimensionality and molecule-surface specific interaction. These effects are of special importance when continuous phase transitions issues are concerned.

### *2.1 Anisotropic fluids and low dimensionality effects*

Some of these issues can be addressed by studying liquid-crystal as model anisotropic fluids confined in uniaxial nanochannels. In order to illustrate this point, 4-n-octyl-4-cyanobiphenyl (8CB) has been chosen as an archetype mesogen. The phases sequence of bulk 8CB on increasing temperature is crystal (K), smectic A (A), nematic (N) and isotropic (I) with the following transition temperatures: $T_{KA}$=294.5 K, $T_{NA}$=306.3 K and $T_{NI}$=313.5 K.

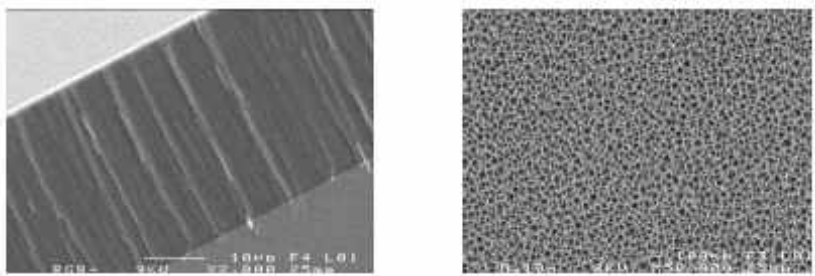

**Figure 1.** Scanning electron micrographs of the porous silicon film. (a) Side view at low magnification showing the 30 μm thick porous layer attached to the silicon substrate and (b) top view at higher magnification.

Two different porous materials have been chosen: porous alumina membranes (anopores) purchased from Whatman Company, and laboratory made porous silicon (PSi), as shown in Fig. 1. Anopore membranes porous geometry is formed by a parallel arrangement of nanochannels of average diameter around 30 nm and length 65 μm. The aspect ratio of this type of porous membranes exceeds 1000:1 and confers a low dimensionality (quasi 1D) to the system. The preferential alignment of all the channels perpendicularly to the membrane surface prevents powder average limitations when measuring anisotropic observables of unidimensional nanoconfined systems. They are therefore model porous materials for studying nanoconfinement effects in a quasi-1D geometry with macroscopic order along the pore axis.
The so-called 'columnar form' of PSi we use has been obtained by electrochemical anodization of heavily p-doped (100)-oriented silicon substrates in an HF electrolyte solution. All the properties of PSi layers, such as porosity, thickness, pore diameter, microstructure are known to be strongly dependent on the anodization conditions and have been extensively studied [10]. Different morphologies are obtained for lightly doped (p-type or n-type) silicon, leading to various degrees of branching and eventually to isotropic microstructures [11]. Conversely, anodization of heavily doped (100)-oriented silicon leads to highly anisotropic porous layers consisting of macroscopically long channels (diameter ~300 Å) running perpendicular to the wafer surface. However, the internal surface of the 1-D pores is characterized by a significant disorder at microscopic lengthscales i.e. ~1 nm.

This additional feature of the columnar form of PSi, as compared to anopores, which present rather smooth surfaces, can be used in order to study specifically the effects of different surface morphologies on the properties of the phase confined in one-dimensional porous channels.

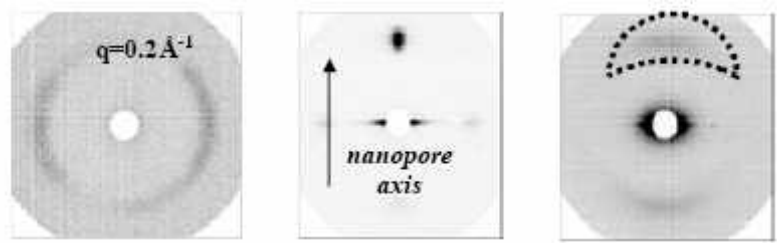

**Figure 2.** Small angle neutron scattering patterns of 8CB in the smectic phase. (a) bulk at 300K. (b) Confined in the channels of porous alumina at 300K. (c) Confined in the channels of porous silicon at 290 K.

A first consequence of manipulating anisotropic fluids in uniaxial nanochannels is illustrated in Fig. 2. Below 306.7 K, bulk 8CB, develops a quasi-long range smectic order. The periodic modulation of the molecular density within the fluid due to the formation of layers can be evidenced by X-rays or neutron diffraction experiments.

The small angle neutron scattering pattern on the 2D detector of bulk 8CB is displayed in Fig. 2 (a). A clear diffraction ring appears for an angle of diffraction corresponding to a value of the momentum transfer $q$ equals to 0.2 Å$^{-1}$. This peak is the signature of the quasi-long range translational order in the smectic phase. The $q$-value corresponds to a modulation of the density across the smectic layers of about 3 nm. In bulk conditions, there is no preferred direction so that the sample splits into a large number of domains with independent smectic order orientations. The formation of a large number of smectic domains with no preferential orientations is proved by the circular symmetry of the diffraction peak.

The small angle neutron scattering pattern of 8CB confined in anopores at 300 K (cf. Fig. 2(b)) reveals a unique Bragg peak at 0.2Å$^{-1}$. This sharp peak shows that the bulk smectic order is maintained in confinement. In addition, it proves that the orientation of the smectic domains is not isotropically distributed as it is for the bulk system but follows precisely the main axis of the nanochannels. This preferential unidirectional orientation induced by anisotropic confinement is extremely well-defined and compares with the bulk one under the application of a strong magnetic field.

When confined in PSi at 290 K, 8CB exhibits two symmetric Bragg peaks having a crescent shape with an angular aperture (half width at half maximum) of 30 degrees. It proves the existence of a preferential orientation of the smectic director along the nanochannels axis, although not as pronounced as for 8CB confined in anopores.

*2.2 Surface morphology and quenched disorder effects*

LC's are elastically soft and, when confined in mesoporous materials, may directly couple to the surface of the solid matrix. The new boundary conditions imposed to the confined LC by the porous matrix can be expressed in terms of an external field. The effect of this external field on the structure, dynamics and phase behavior of the confined phase is therefore directly dependent on the morphology of the porous solid. As an example, Bellini *et al.* have shown that confinement in random porous materials (aerogels) could be used in order to introduce random fields, which couple to the LC order parameters [12]. Under these conditions of confinement, the weakly first order isotropic-nematic and the continuous nematic-smectic phase transitions are shown to be strongly affected. The most striking observation is the absence of a true nematic-to-smectic transition, which is replaced by the gradual occurrence of a short-range ordered phase. This confirms the general theoretical prediction that a transition breaking a continuous symmetry is unstable toward arbitrarily small quenched random fields.

Quenched disorder effects have been generally expected to be small or absent for fluids confined in channels, since they correspond to regular geometries. Indeed, moderate changes in the structure and phase transitions of LC's confined in one-dimensional porous materials such as nanoporous alumina (anopore) or track-etched (nuclepore) membranes have been essentially explained in terms of surface anchoring and finite size with negligible quenched disorder effects.

We have recently demonstrated that the corrugated inner surface of PSi nanochannels actually acts as a random field that couples to the order parameter of the interfacial fluid, producing strong quenched disorder effects for a LC confined in a one-dimensional porous material [13,14,15]. In addition to usual surface and finite size effects, quenched disorder introduced by PSi leads to a progressive and moderate increase of the smectic correlation length from molecular size to about 150 Å far below the temperature domain where macroscopic transitions are encountered in the bulk fluid. From a theoretical point of view, such a temperature variation and the value of the saturating correlation length has been actually expressed in terms of a competition between the strength of the random field and the elasticity of the confined phase only in the limit of very weak and homogeneous disorder. This experimental occurrence of quenched disorder effects in a regime of anisotropic and very strong density of disorder offers promising openings for experiments and theory.

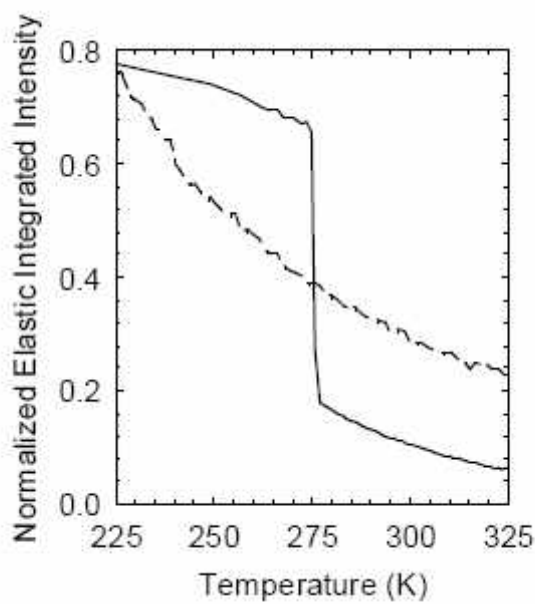

**Figure 3.** Incoherent neutron fixed window elastic scans of bulk 8CB (solid line) and confined in PSi nanochannels (dashed line) during cooling. The intensity is integrated over a $q$-range from 0.4 to 1.9 Å$^{-1}$ and normalized at the lowest temperature (30 K).

Changes in the structure and phase behavior of 8CB confined in PSi channels are going along with drastic changes of the molecular mobility. This phenomenon is illustrated in Fig. 3, which displays the incoherent elastic intensity measured on cooling with a high-resolution neutron backscattering spectrometer IN16 (ILL, Grenoble). The use of fully hydrogenated samples allows one to measure mainly the incoherent scattering processes, hence probing individual atomic motions. The incoherent elastic intensity reflects the fraction of the molecular relaxation processes, which occurs within the frequency resolution of the spectrometer, i.e. 200 MHz (FWHM).

This quantity is small for the bulk liquid at 325 K, as expected for low-viscous fluids when fast translational diffusion occurs within the spectrometer timescale. A gradual increase of the elastic intensity occurs on cooling, as a consequence of the temperature dependence of the fluid viscosity. A sharp increase is observed at about 275 K, which corresponds to the bulk crystallization temperature and subsequent arrest of most of the molecular degrees of freedom.

The elastic intensity of 8CB confined PSi is surprisingly large at 325K, where the system is fully liquid. This result implies that the translational diffusion of the nanoconfined liquid is strongly reduced in confinement. This effect should not be associated to a direct geometric confinement effect, since the molecular mobility probed by this technique corresponds to molecular displacements, which are smaller than the pore size. Most probably, it results from surface interaction effects, which tends to reduce the molecular mobility of the interfacial liquid. Given the large surface-to-volume ratio of the nanophase, the hindered dynamics induced at the boundary conditions may finally affects the overall dynamics of the confined phase. On decreasing temperature, the elastic intensity continuously increases until it reaches the level of the bulk crystal at 225 K. This continuous slowdown of the molecular dynamics of nanoconfined 8CB compares somehow with the behavior of glassy systems and contrasts with the bulk one. It reflects the different phase behavior of 8CB in silicon nanopores, which does not crystallize in this temperature range but presents a continuous increase of short-range smectic correlations induced by the high corrugation of the channel surface.

*2.3 H-bonding surface interactions and bioprotectant solutions*

Previous section underlies the fundamental importance of the surface morphology, in terms of quenched disorder effects. The nature of the interfacial interaction energy is crucial, especially in the case of strongly interacting fluids, such as H-bonding systems. This situation is encountered for an important class of systems, which are aqueous and alcohol solutions. These fluids are widely used and relevant to different domains of technological or biological interest. One of these concerns biopreservation. Highly concentrated bioprotectant solutions, usually prepared with ammonium sulfate, glycerol and disaccharides, are widely used in order to maintain proteins in their native state. Indeed, the stabilization of bio-molecules such as proteins is a prerequisite, to possible manipulations and use in vitro and in biotechnological devices.

From a fundamental point of view, a new level of complexity is awaited for confined bioprotectant solutions, which are multi-component systems with strong and selective H-bond interactions. They are also expected to display concentration fluctuations or even nano-phase separations, which may add to structural and dynamical heterogeneities associated with nanoconfinement.

The details of the physico-chemical mechanism of bioprotection at the molecular level remain obscure, despite unambiguous demonstrations of the cryopreservative and lyopreservative skills of disaccharide or polyalcohol solutions [16,17]. A general agreement concerns the crucial role played by the coupling between the associating character of the liquid solutions (intermolecular structuring carried by hydrogen bonds) and peculiar dynamical properties (most bioprotective solutions are glass-forming systems).

Understanding how confinement affects both the structural properties and the molecular dynamics of biopreservative solutions would provide basic information, which are necessary to model the influence of the environment that is usually encountered *in vivo*. It includes issues related to nanoconfinement effects within intermembranar spaces or specific interactions of constituents with interfacial functional groups.

Glycerol/Trehalose binary solutions have been recently highlighted by experiments and simulations for its optimal biopreservation properties, in particular in terms of stabilization of proteins in native configuration [17]. The two components are glass-forming liquids with very different glass transition temperatures, which lead to a large variation of the dynamical properties with concentration (strong departure from the Gordon-Taylor law).

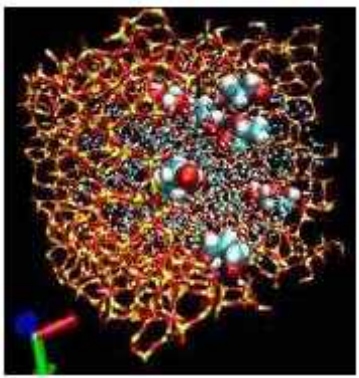

**Figure 4.** Glycerol bioprotectant solution confined in a silica nanochannel. Snapshot of a molecular dynamics simulation.

A study of the dynamics of bulk and confined solutions has been performed by solid state NMR. Fig. 5 displays NMR proton spin-lattice relaxation times measured at 300 MHz on bulk glycerol, bulk glycerol/trehalose (40:60 w%) and glycerol confined in 300Å anopores. At this frequency, the $T_1$ curve has a minimum at a temperature that strongly shifts, depending on concentration or confinement. It can be safely assigned to the primary $\alpha$ relaxation of the fluid, while the mixture exhibits a second relaxation at high temperature, which can be tentatively associated to trehalose. NMR unambiguously reveals antagonist effects of trehalose concentration and nanoconfinement on the glycerol molecular dynamics at the nanosecond time scale. However, it lacks the spatial and spectral analysis that can be provided by other techniques like molecular dynamics simulation (MD) or incoherent quasielastic neutron scattering (QENS).

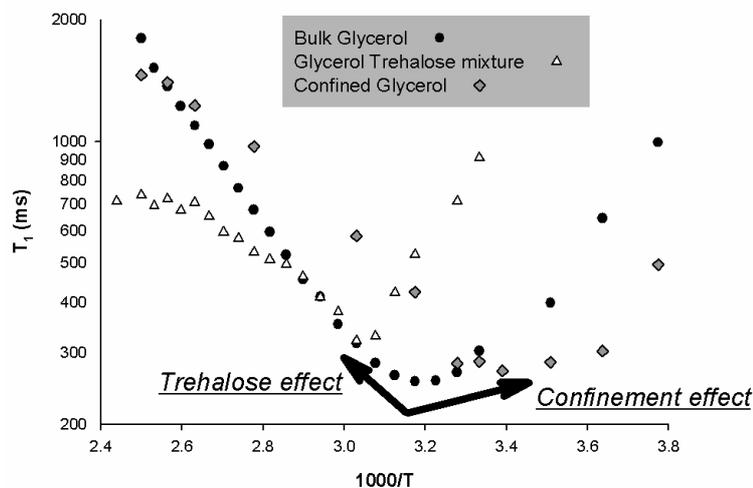

**Figure 5.** Effects of confinement and trehalose amount on glycerol dynamics by T1 NMR experiments

Confined effects on glycerol solution are currently investigated by MD molecular simulations (cf. Fig. 4). This approach relies on the creation of a realistic model of a mesoporous silicate, according to a procedure recently used for liquid methanol [18]. It will allow producing a realistic description of the interfacial interactions between the fluid and the matrix in terms of surface silanol groups and irregular inner surface of the nanochannel.

## 3  Ultra-confinement effects in self-assembled nanoporous crystals

Self-assembly underlies the formation of a wide variety of biological and supramolecular nanotubular structures. The versatility of the structrures results from the weak bonding due to non-covalent interactions [19,20]. In most cases, self-assembled compounds are constructed from two interpenetrating sublattices. Short peptides and other organic molecules yield numerous examples of molecular self-assembled systems, defining host and guest sub-systems [20,21]. Nano-peapods of fullerene arrays inside single wall carbon nanotubes constitute a new type of self-assembled hybrid structures [22]. The common feature of all these materials is to present the ultimate one-dimensional confinement for the guest molecules. Aperiodicity in these composite materials may appear rather naturally due to the misfit of host and guest parameters along the channels or the tubes. The fundamental feature of aperiodic crystals concerns their structures, which can be understood in the frame of superspace crystallography [23]. These crystals may show low lying excitations generated by the infinite degeneracy of their ground state in relation to this possibility of describing these structures in a higher dimensional space [23, 24]. Among the various types of incommensurate systems one of the simplest is illustrated by uniaxial composite crystals. Incommensurate composite structures can be considered as a new state of matter like liquid or solid states. The hydrodynamics of these materials has been the object of several theoretical papers [24,25,26].

Urea is the smallest molecule containing the peptide linkage and urea inclusion compounds constitute a very simple paradigm example of self-assembled nanoporous crystals [27,28,29]. Alkane-urea crystals are needles with perfectly defined and aligned channels. These channels have very huge aspect ratio around 20 000 000. They are opened at both ends, as nicely proved by molecular diffusion observation through space resolved micro-Raman scattering measurements [30]. In urea inclusion crystals, the host network is made of honeycomb-like channels formed by helical ribbons of urea molecules. Channels are almost cylindrical with an available diameter of 5.5 Å, which can accommodate linear guest molecules like alkanes ($C_nH_{2n+2}$) [27,29]. At room temperature, guest molecules occupy a six-fold symmetry site which results from an orientational disorder about the long molecular axis. Aperiodicity appears along the channels. The stoichiometry is determined by the pitch of the urea helices (here $c_{urea}$=11.02 Å at 300K) and the guest length (here $c_{alkane}$ which parameter is mainly a linear function of the number n of carbon atoms). In such composite crystals, a misfit parameter α is defined by the ratio of the host and guest periodicities. The whole crystal is the union of two incommensurate modulated lattices, and the diffraction pattern is not the simple superimposition of the two uncorrelated ones but reveals new intensity relations and additional reflections. In the present case, a four dimensional superspace description gives the positions of the whole set of Bragg :

$$\vec{q}_{hklm} = h\vec{a}^* + k\vec{b}^* + k\vec{c}_h^* + m\vec{c}_g^*$$

$\vec{a}^*$, $\vec{b}^*$, $\vec{c}_h^*$ and $\vec{c}_g^*$ are the conventional reciprocal vectors, h and g indexes define host and guest respectively.

The different Bragg structure reflections of the diffraction pattern can be classified as follows : (h, k, 0, 0) are common reflections from the basal periodic plane, (h, k, l, 0) and (h, k, 0, m) render account of the mean periodicity of respectively the host and the guest sub-lattices, (h, k, l, m) with l and m different from zero are intermodulation satellites. This characteristic diffraction signature was reported in urea-alkane crystals [31].

### 3.1 Free sliding and selective compressibility

Quite new properties, such as a molecularly selective capillary [30] and selective compressibility [32], where recently reported in these materials. These remarkable features rely on the aperiodicity of the structures, which theoretically enables an homogeneous displacement (or compression) of one sublattice with respect to the other one without any restoring force. This property results from the infinitely degenerate ground state of an infinite aperiodic structure.

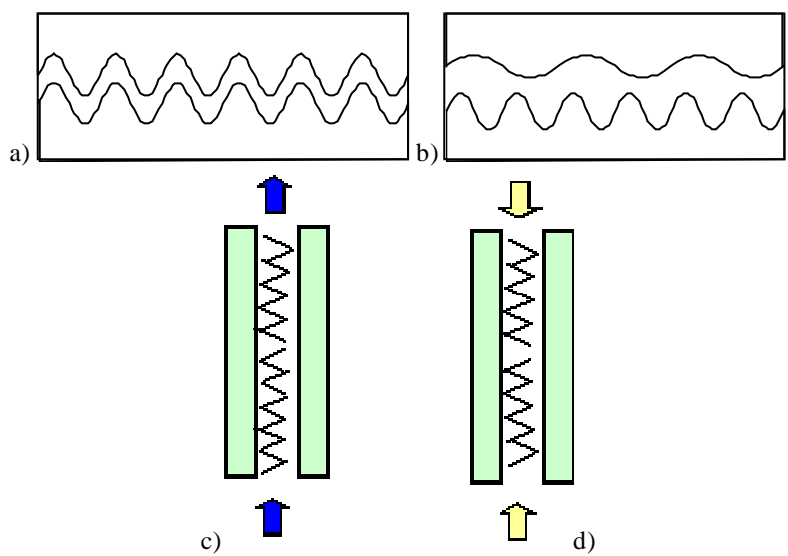

**Figure 6**. a) commensurate lock-in as in usual crystals; b) illustration of the free sliding allowed by aperiodicity; c) illustration of molecularly selective capillary through osmotic forces [30], d) illustration of a molecular press [32].

Diffraction techniques were used to study the structural modifications under pressure. Neutron scattering experiments were performed at the Laboratoire Léon Brillouin (Orphée reactor, Saclay, France). The data were collected on a triple axis spectrometer installed on a cold neutron source (4F). The incident wave vector was $k_i = 1.55$ Å$^{-1}$ with a refrigerated beryllium filter removing higher order contaminations. Hydrostatic pressure up to 6 kbar was obtained in a helium gas pressure cell.

The retained guest compounds are nonadecane ($C_{19}H_{40}$) and hexadecane ($C_{16}H_{34}$) which give the respective misfit parameter $\alpha=0.418$ and $\alpha=0.487$ at ambient condition. The single crystals were prepared by slow evaporation of mixed solution of urea and alkane in a mixture of ethanol and isopropyl alcohol. The hexadecane-urea crystal is fully hydrogenated. However, the very high spatial resolution of the triple axis spectrometer used with very cold neutrons makes the integrated incoherent neutron scattering quite small compared to the Bragg peaks.

These measurements were performed in the hexagonal high symmetry phase, at room temperature up to a pressure of 6 kbar. From these data, the pressure evolution of the misfit parameter is extracted as shown in Fig. 7. Above 1 kbar, we observe the selective compressibility of both host and guest sublattices through the pressure variation of the misfit parameter. This observation is similar in nonadecane-urea where we demonstrated that the data are the same for upstroke and down stroke. In hexadecane-urea, when the misfit parameter meets the ½ commensurate value, it presents a plateau on the pressure range of 800 bar. There, the system behaves as a usual crystal. This is the evidence of a commensurate lock-in inside this nanoporous crystal. At higher pressure, the crystal presents again a depinning with the same relative slope than at low pressures. Decreasing pressure from 6 kbar to 4.2 kbar, on the plateau shows that there is no hysteresis.

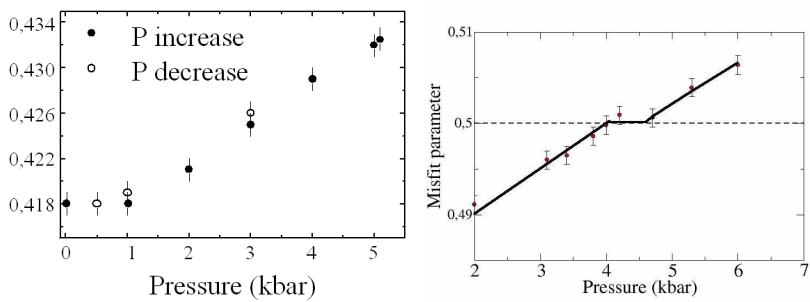

**Figure 7.** left) Selective compressibility of the nanoporous nonadecane-urea crystal; right) selective compressibility and pressure induce lock-in in hexadecane-urea.

The realization of such a molecular press is of fundamental importance since it allows a continuous and controlled tuning of molecular properties under extreme one dimensional confinement [33]. The versatility of guest molecules open a very broad field of research concerning manipulation of functionalized molecules [34].

*3.2 Collective dynamics in aperiodic compounds*

Because of the lack of translational symmetry, the concept of Brillouin zone vanishes and the collective molecular motions are no longer characterized by a wave-vector inside the Brillouin zone. The dynamics of aperiodic crystals are then totally different from the one of periodic crystals. Due to the two length scales, there are twice as many hydrodynamic variables as in a conventional crystal but these additional variables cannot be interpreted as real displacements of any of the sub-lattices [24,25,26]. If the interaction between the sub-lattices could be neglected, on would get two zero energy, each ones corresponding to a rigid displacement of one sub-lattice with respect to its fixed counterpart.
In real materials, this condition is not fulfilled since the crystal would not be stable. In previous experiments, several direct evidence of a rather important interaction between the different sub-lattices were reported through the observation of intermodulation satellites [31] or pressure induced lock-in [33]. Interaction between the two sublattices may couple acoustic modes to give rise to two new modes:
-an actual unique acoustic mode for the composite
-another mode (second sound) corresponding to the relative anti-translation of both sublattices. This sliding appears in theory at zero frequency since the relative shift of the lattices does not cost any potential energy due to incommensurability. In a superspace description, this displacement correspond to a change of the phase of the incommensurate intermodulation perpendicular to the physical space [23,24]
Due to aperiodicity, Bragg peaks are not expected to be a center of zone for the crystal. Both dispersion and damping are strongly affected by the lack of translational symmetry in the crystal. This accounts for the fact that, away from the unique real Brillouin zone center Q= (0 0 0 0), pure acoustic features are never found, even close to the main strong Bragg peaks. Since, it is proved that lock-in energy terms are important in these self-assembled materials, the urea sublattice is expected to strongly feels the nature of the guest sub-lattice motions. Then, according to the relative positions of the host and guest

Braggs along this incommensurate direction, different coupling and consequently damping and gaps could be expected at other strong Braggs.

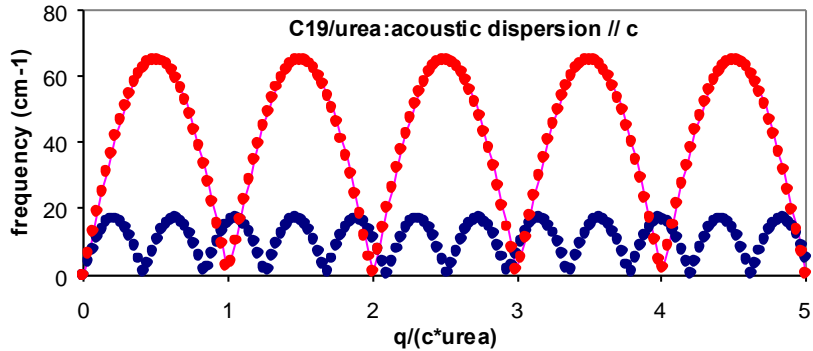

**Figure 8.** Collective longitudinal acoustic modes along the incommensurate direction of an aperiodic crystal

Two previous works were previously performed using Brillouin scattering around the Brillouin zone center Q= (0 0 0 0). They came to two contradictory results. A first study reported an underdamped supplementary mode in hexadecane-urea [35],a mode assigned to the sliding mode. This observation was not reproduced in a second Brillouin measurement study [36].These measurement however reveals strong quasi-elastic components whose origin, for one of them could be compatible with an overdamped sliding mode. The frequency of the LA mode gives the sound velocity of the composite. Its value, as measured by Brillouin scattering, is associated to the whole urea-alkane and was found equal to 4.95 km/s. The damping of the LA mode as measured by Brillouin scattering propagating along c* also appears anomaly large. Several explanations may be proposed lying either in the disorder of the guest phase or in the coupling at non-finite wavevector of the sound mode and the sliding mode. A similar anomalous damping was also observed in the neutron inelastic scattering measurements. Previously, by using the results of the molecular press, the inter alkane force constant was determined [36] allowing to compute the sound velocity in the uncoupled alkane sublattice $V_a = c_a \sqrt{(k/M_a)}$ = 3.96 km/s ($C_{19}H_{40}$). The longitudinal acoustic dispersion curves of both uncoupled sublattices are represented in Fig. 8 where the upper curve, related to the urea one, is computed from the true composite acoustic sound velocity, which has an intermediate value between the ones associated to each sub-lattice. The lower curve is drawn in agreement with the above-mentioned $V_a$.

Further features may happen in the reciprocal space where the dispersion branches associated to both host and guest sub-lattices are expected to cross. This should lead to a repulsion of the crossing branches with formation of a rich hierarchical set of gaps when $\omega_h(Nq) = \omega_g(Mq)$. In our measurement, no such gap is evidenced but one can notice that a faster variation of the anomalous damping appears as shown in the figure where N=4 and M=10. A theoretical modelisation including actual interaction energy terms should be necessary to go further on.

Previous neutron studies were reported in incommensurate composites (mercury chains in ASF6). In these cases, different sound velocities were reported depending on the nature of the retained Bragg peaks. Aperiodicity in this non-stochiometric compound provided a nice example of pure one-dimensional liquid in a solid state matrix. The synthesis of the numerous work concerning structure and dynamics in matter where aperiodicity drastically increases the number of degrees of freedom should be very fruitful.

## 4  Conclusions

Organic phases confined in containers of nanometer size exhibit very exceptional properties, which depart from the one of the same system filling a glass of macroscopic dimension. They show anomalous molecular mobility, peculiar phase transitions and unusual intermolecular ordering. These phenomena depend on the size of the nanocontainer, on the nature and the strength of its interaction with the fluid.
In the present contribution, we have focussed on the importance of the surface morphology, in terms roughness when considered at the level of a few nanometers and atomic ordering at the molecular scale. The introduction of new concepts in the field of nanofluidics such as quenched disorder or aperiodicity is illustrated by experimental studies carried out on a variety of molecular materials including fluids in silica or silicon mesopores and self-assembled organic phases.

We show that strong irregularities of the surface of nanochannels may couple to the structural order parameters of the confined phase. This feature has drastic consequences on the stability of some usual condensed phases, such as mesomorphic fluids, which are very sensitive to quenched disorder effects. The observed short-range-ordered confined phases, display a strongly hindered molecular dynamics with a 'glass-like' slowdown on decreasing temperature.

Going down to ultra-confinement conditions, reached by the channels of intergrowth nanoporous compounds, we draw the surprising conclusion that new hidden degrees of freedom, provided by incommensurability, can lead to a recovery of large translation molecular mobility.

### Acknowledgements

The financial supports from the *Centre de Compétence C'Nano Nord-Ouest* and *Rennes Metropole*. are expressly acknowledged. Some of the reported results are parts of the PhD thesis of R. Busselez, who benefits from a doctoral research grant from the *Brittany Region*.
Beam time allocation in neutron scattering large scale facilities (Institut Laue Langevin, Grenoble and Laboratoire Léon Brillouin, CEA-Saclay) are acknowledged.